\documentclass[%
aip,%
amsmath,amssymb,floatfix,
reprint,%
twocolumn,%
longbibliography,
reprint,%
]{revtex4-1} 

\allowdisplaybreaks

\usepackage{amsmath,amsthm,latexsym,amssymb,amsfonts,epsfig}

\usepackage[english]{babel}
\usepackage[utf8]{inputenc}			
\usepackage{graphicx} 
\usepackage{color}
\usepackage{bm}

\usepackage{mathtools}
\usepackage{mathrsfs} 

\usepackage[left= 1.5cm, right = 1.5cm, top= 2cm, bottom = 2cm]{geometry}  

\usepackage{enumerate}

\usepackage[colorlinks=true,citecolor=OliveGreen, linkcolor=blue]{hyperref}

\usepackage{color}
\usepackage[svgnames,dvipsnames]{xcolor}

\usepackage{amsmath}
\usepackage{amsfonts}
\usepackage{amssymb}

\usepackage{epstopdf}

\usepackage{type1ec}

\usepackage{placeins}
\usepackage[titletoc,title]{appendix}

\newcommand{\vect}[1]{\boldsymbol{#1}}

\newcommand{\R}{\vect{r}}

\newcommand{\Intd}{\mathrm{d}}

\definecolor{AirForceBlue}{rgb}{0., 0.34, 0.66}

\emergencystretch=\maxdimen
\begin{document}

\title{Steady azimuthal flow field induced by a rotating sphere near a rigid disk or inside a gap between two coaxially positioned rigid disks}

\date{\today}

\author{Abdallah Daddi-Moussa-Ider}
\email{ider@ds.mpg.de}
\affiliation{Institut f\"{u}r Theoretische Physik II: Weiche Materie, Heinrich-Heine-Universit\"{a}t D\"{u}sseldorf, Universitätsstraße 1, D--40225 D\"{u}sseldorf, Germany}

\affiliation{Abteilung Physik lebender Materie, Max-Planck-Institut für Dynamik und Selbstorganisation, Am Faßberg 17, D--37077 Göttingen, Germany}

\author{Alexander R.\ Sprenger}
\affiliation{Institut f\"{u}r Theoretische Physik II: Weiche Materie, Heinrich-Heine-Universit\"{a}t D\"{u}sseldorf, Universitätsstraße 1, D--40225 D\"{u}sseldorf, Germany}

\author{Thomas Richter}
\affiliation{Institut für Analysis und Numerik, Otto-von-Guericke-Universität Magdeburg, Universitätsplatz 2, D--39106 Magdeburg,Germany}

\author{Hartmut Löwen}
\affiliation{Institut f\"{u}r Theoretische Physik II: Weiche Materie, Heinrich-Heine-Universit\"{a}t D\"{u}sseldorf, Universitätsstraße 1, D--40225 D\"{u}sseldorf, Germany}

\author{Andreas M.\ Menzel}
\affiliation{Institut für Physik, Otto-von-Guericke-Universität Magdeburg, Universitätsplatz 2, D--39106 Magdeburg, Germany}

\begin{abstract}
	
	Geometric confinements play an important role in many physical and biological processes and significantly affect the rheology and behavior of colloidal suspensions at low Reynolds numbers.
	On the basis of the linear Stokes equations, we investigate theoretically and computationally the viscous azimuthal flow induced by the slow rotation of a small spherical particle located in the vicinity of a rigid no-slip disk or inside a gap between two coaxially positioned rigid no-slip disks of the same radius.
	We formulate the solution of the hydrodynamic problem as a mixed-boundary-value problem in the whole fluid domain, which we subsequently transform into a system of dual integral equations.
	Near a stationary disk, we show that the resulting integral equation can be reduced into an elementary Abel integral equation that admits a unique analytical solution.
	Between two coaxially positioned stationary disks, we demonstrate that the flow problem can be transformed into a system of two Fredholm integral equations of the first kind. 
	The latter are solved by means of numerical approaches.
	Using our solution, we further investigate the effect of the disks on the slow rotational motion of a colloidal particle and provide expressions of the hydrodynamic mobility as a function of the system geometry.
	We compare our results with corresponding finite-element simulations and observe very good agreement.
	
\end{abstract}

\maketitle

\section{Introduction}

Hydrodynamic interactions in viscous flows are ubiquitous in nature and find numerous applications in various industrial and environmental processes.
Simultaneously, confinements play a pivotal role in a wide range of biological and biotechnological processes, including the dynamics of polymer solutions and melts in microfluidic devices~\cite{zhou2019effect, muller2009hydrodynamic, piette2019hydrodynamic, kanso2019order}, DNA translocation through pores~\cite{storm2005fast, izmitli2008effect}, transport and rheology of red blood cell suspensions in microcirculation~\cite{secomb86, pozrikidis05axi, mcwhirter09, Freund_2014, secomb17, barthes16}, colloidal gelation~\cite{furukawa2010key}, biofilm formation in microchannels~\cite{rusconi10, rusconi11, drescher13, kim14}, 
and swimming behavior of active self-propelled agents in viscous media~\cite{berke08, yazdi17, bianchi2017holographic, reigh17, bianchi2017holographic, daddi18, dhar2020hydrodynamics}.

Fluid flows at small length scales are characterized by low Reynolds numbers, where viscous forces typically dominate inertial forces.
Under such conditions, the fluid dynamics can well be described by the linear Stokes equations~\cite{kim13}.
Over the past few decades, there has been mounting interest in the theoretical and experimental characterization of the behavior of hydrodynamically interacting particles near confining interfaces~\cite{diamant09}.
These include, for instance, a flat rigid wall~\cite{mackay61, gotoh82, cichocki98, lauga05, swan07, franosch09, felderhof12, decorato15, huang15, rallabandi17obstacle}, a planar surface with partial slip~\cite{lauga05}, a flat interface separating two immiscible fluids~\cite{lee79, berdan81, blawz10theory, blawz10}, an interface covered with a surfactant~\cite{blawz99a, blawz99b}, a rough boundary characterized by random surface textures~\cite{kurzthaler2020particle}, or a soft deformable membrane possessing elastic and bending properties~\cite{felderhof06, bickel06, bickel07, boatwright14, daddi16, daddi16c, junger15, salez15, daddi17,daddi17pof, daddi18epje, rallabandi18, daddi18stone, daddi18coupling, hoell19creeping}.
Thanks to the advent of new particle tracking and measurement techniques, the field has benefited from important recent advances in the characterization of the behavior of colloidal particles near confinement at small scales~\cite{lobry96, graham11, faucheux94, lin00, traenkle16}.

From a chronological standpoint, one of the first attempts to address the creeping flow induced by a spherical particle confined between two infinitely extended planar walls dates back to Fax\'{e}n~\cite{faxen21}.
One century ago, Fax\'{e}n provided in his doctoral dissertation a few approximate analytical expressions of the hydrodynamic mobility function for parallel translational motion in a channel bounded by two flat plates.
Later, using the method  of images, Liron and Mochon~\cite{liron76} obtained in a pioneering work an exact solution of the Stokes flow induced by a point-force singularity acting between two parallel no-slip walls.
The problem of fluid motion in a channel bounded by two no-slip walls has further been addressed using the multipole expansion technique~\cite{bhattacharya02, swan10} and a strong interaction theory~\cite{ganatos80a, ganatos80b}.

In the present article, we proceed a step further by examining the low-Reynolds-number flow induced by a point-like particle rotating near a rigid finite-sized no-slip disk or between two coaxially positioned rigid finite-sized no-slip disks of the same radius.
Mathematically, we model the situation using a rotlet (also called a point-torque or point-couple) singularity acting on the surrounding fluid medium.
We formulate the flow problem at hand as a mixed-boundary-value problem which we subsequently transform into a system of dual integral equations on the domain boundary.
For their solution, we employ conventional procedures outlined by Sneddon and Copson so as to express the solutions of the flow problems in terms of convergent definite integrals.
Moreover, we quantify the effect of the confining finite-sized disks on the rotational motion by calculating the effect on the corresponding hydrodynamic mobility function.

The remainder of this article is organized as follows.
In Sec.~\ref{sec:singleDisk} we derive the solution of the hydrodynamic equations for a rotlet singularity acting near a fixed no-slip disk.
We show that the induced velocity field can be presented in a compact analytical form in terms of a definite one-dimensional integral.
Afterwards, we obtain in Sec.~\ref{sec:twoDisks} a semi-analytical solution of the flow problem inside a gap between two coaxially positioned rigid no-slip disks.
We demonstrate that the solution can be reduced into a system of two Fredholm equations of the first kind that can be solved by means of standard numerical approaches.
Finally, concluding remarks are contained in Sec.~\ref{sec:conclusions}.
Technical aspects and simulation details regarding the finite-element method we employ to compare our theory with are shifted to the appendices.

\section{Solution near a single disk}
\label{sec:singleDisk}

\subsection{Problem formulation}

\begin{figure}
	\centering
	\includegraphics[scale=1.33]{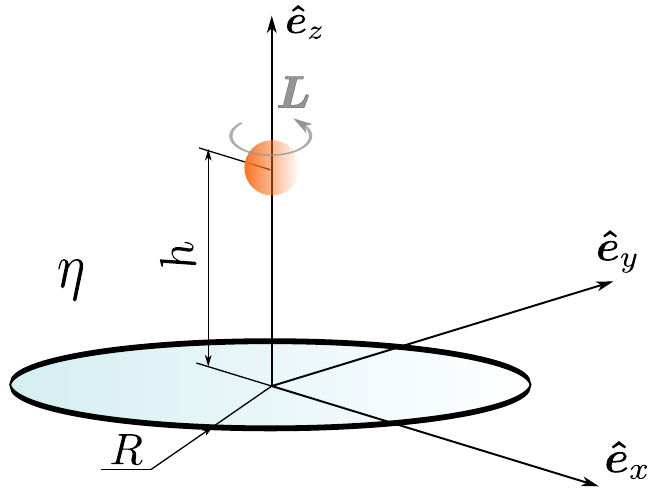}
	\caption{(Color online) Graphical illustration of the system setup. 
	A point-like particle undergoing slow rotational motion near a rigid no-slip disk of radius~$R$ sets the fluid into motion.
	The center of the particle is located at a distance~$h$ above the center of the disk, while the surrounding viscous fluid medium is characterized by a dynamic shear viscosity~$\eta$. 
	$\vect{L} = L \vect{\hat{e}}_z$ sets the torque acting via the particle at the particle position on the fluid.
	}
	\label{IllusSingle}
\end{figure}

First, we examine the low-Reynolds-number dynamics of a point-like particle undergoing rotational motion near one fixed finite-sized disk of radius~$R$.
We assume a no-slip boundary condition to hold at the surface of the disk.
In addition, we suppose that the disk is located within the plane~$z=0$ and that the center of the disk coincides with the origin of our coordinate frame; see Fig.~\ref{IllusSingle} for an illustration of the system setup.
In addition, we assume that the fluid is incompressible and Newtonian with constant shear viscosity~$\eta$.

At low Reynolds numbers, the fluid dynamics is thus governed by the steady Stokes equations~\cite{happel12}
\begin{equation}
	\eta \boldsymbol{\nabla}^2 \vect{v} - \boldsymbol{\nabla} p + \vect{F}_\mathrm{B} 
	 = \vect{0} \, , \qquad \boldsymbol{\nabla} \cdot \vect{v} = 0\, ,    
\end{equation}
wherein $\vect{v}$ and~$p$ denote the hydrodynamic velocity and pressure fields, respectively.
In addition, $\vect{F}_\mathrm{B}$ represents an arbitrary bulk force density acting on the fluid at position~$\R_0 = h \vect{e}_z$ with~$\vect{e}_z$ denoting the unit vector directed along the $z$~direction. 
The torque $\vect{L}$ on the particle is transmitted to the fluid and linked to the surface force density $\vect{F}$ acting on the fluid via the surface of the spherical particle
\begin{equation}
	\vect{L} = \oint_A \left( \R - \R_0 \right) \times \vect{F} \, \Intd S \, , 
\end{equation}
with~$A$ denoting the surface area of the tiny particle.
In the point-particle approximation, the asymmetric dipolar term in the multipole expansion is associated with the flow field induced by a rotlet singularity of strength~$\vect{L}$ acting above the disk at position~$\R_0$.
Here we consider the case in which the point torque is directed along the axis of symmetry of the disk and set $\vect{L} = L \vect{e}_z$.

´

In an unbounded (infinite) fluid medium, the flow field induced by a rotlet singularity is given by
\begin{equation}
	\vect{v}^\infty (\vect{r}) = \frac{1}{8\pi\eta} 
	\frac{\vect{L} \times \vect{s}}{s^3} \, ,  
\end{equation}
wherein~$\vect{s} = \vect{r} - \vect{r}_0$ and $s = |\vect{s}|$ is the distance from the singularity position.
Using cylindrical coordinates $(r,\phi,z)$, the azimuthal component of the flow velocity field induced by a free-space rotlet oriented along the $z$ direction reads
\begin{equation}
	v_\phi^\infty (r, z) = \frac{K r}{\left( r^2 + \left (z-h \right)^2 \right)^{\frac{3}{2}}} \, ,  \label{vInf}
\end{equation}
where we have defined, for convenience, the abbreviation $ K = L / \left( 8\pi\eta \right) $ of dimension (length)$^3$(time)$^{-1}$.

The solution of the flow problem in the presence of the confining disk can generally be expressed as a linear superposition of the solution in an unbounded fluid medium, given by Eq.~\eqref{vInf}, and a complementary solution that is required to satisfy the underlying regularity and boundary conditions for the total induced flow field.
Specifically,
\begin{equation}
	v_\phi = v_\phi^\infty + v_\phi^* \, , \label{totalFlowField}
\end{equation}
with~$v_\phi^*$ standing for the complementary solution for the azimuthal flow velocity, also sometimes called the image solution~\cite{blake71}.

The solution of the homogeneous equations governing the fluid motion can be expressed in our case in terms of a single harmonic function as $v_\phi^* = -\partial \Omega / \partial r$ 
(c.f.\ Ref.~\onlinecite[Eq.~(3--3.58)]{happel12}, or Ref.~\onlinecite[Eq.~(5)]{Shail1987}, and references therein for the expression of the complete solution)
with~$\Omega$ satisfying the Laplace equation $\boldsymbol{\nabla}^2 \Omega = 0$.
Then, the harmonic function~$\Omega$ can be written in the general form in terms of a Fourier-Bessel integral of the form
\begin{equation}
	\Omega (r, z) = K \int_0^\infty \omega(\lambda) J_0(\lambda r) e^{-\lambda |z|} \, \Intd \lambda \, ,
\end{equation}
wherein~$\omega(\lambda)$ is an unknown wavenumber-dependent function to be subsequently determined from the prescribed boundary conditions.
In addition, $J_\nu$ denotes the Bessel function~\cite{abramowitz72} of the first kind of order~$\nu$.
The image solution for the azimuthal component of the velocity field is then obtained as
\begin{equation}
	v_\phi^* (r, z) = K \int_0^\infty \lambda \omega(\lambda) J_1(\lambda r) e^{-\lambda |z|} \, \Intd \lambda \, . \label{vIm}
\end{equation}
Evidently, the solution form given by Eq.~\eqref{vIm} satisfies the natural continuity of the azimuthal velocity field at the plane $z = 0$.
We note that the rotlet singularity does not induce a pressure gradient and that the radial and axial components of the fluid velocity vanish.
Therefore, the solution of the flow problem reduces to the search for the azimuthal component of the velocity field only.

\subsection{Boundary conditions and dual integral equations}

We require no-slip boundary conditions on the surface of the disk and assume the continuity of the azimuthal component of the normal stress vector on the plane $z=0$ outside the disk.
Specifically,
\begin{subequations} \label{BCs}
	\begin{align}
		\left. v_\phi^\infty + v_\phi^* \right|_{z=0} &= 0 \quad \text{for} \quad r < R \, , \\[5pt]
		\left. \eta \,  \frac{\partial v_\phi^*}{\partial z} \right|_{z=0^+} 
		- \left. \eta \,  \frac{\partial v_\phi^*}{\partial z} \right|_{z=0^-} &= 0 \quad \text{for} \quad r > R \, .
	\end{align}
\end{subequations}

By inserting the expressions of the free-space and image fields given by Eqs.~\eqref{vInf} and~\eqref{vIm}, respectively, into Eqs.~\eqref{BCs}, we obtain the mixed-boundary-value problem on the inner and outer domain boundaries.
Specifically,
\begin{subequations}\label{dualIntEqsSingle}
	\begin{align}
		\int_0^\infty \lambda \omega(\lambda) J_1(\lambda r) \, \Intd \lambda &= f(r) \quad\quad (r<R) \, , \label{innerProblem} \\
		\int_0^\infty \lambda^2 \omega(\lambda) J_1(\lambda r) \, \Intd \lambda &= 0 \qquad\quad\,\, (r>R) \, , \label{outerProblem}
	\end{align}
\end{subequations}
with the radial function
\begin{equation}
	f(r) = -\frac{r}{\left( r^2 + h^2 \right)^{\frac{3}{2}}} \, , \label{radialFct}
\end{equation}
stemming from the free-space rotlet field.

The solution of the type of dual integral equations stated by Eqs.~\eqref{dualIntEqsSingle} can generally be obtained using the theory of Mellin transforms~\cite{tranter51, titchmarsh48book}. 
We will follow in the present article a different route based on the analytical approach outlined by Sneddon~\cite{sneddon60} and Copson~\cite{copson61}.
In the sequel, we will show that the present dual integral equations problem with Bessel function kernels can conveniently be reduced to an elementary Abel integral equation that may readily be inverted.
This solution strategy has previously been employed to examine the low-Reynolds-number flow induced by nonrotational force singularities near a finite-sized elastic disk possessing shear and bending properties~\cite{daddi19jpsj, daddi2020asymmetric}, the flow field near a no-slip disk~\cite{daddi2020dynamics, kim83}, or the axisymmetric flow due to a Stokeslet acting between two coaxially positioned rigid no-slip disks~\cite{daddi2020axisymmetric}.

We search a solution for the unknown wavenumber-dependent function~$\omega(\lambda)$ of the integral form
\begin{equation}
	\omega(\lambda) = \lambda^{-\frac{1}{2}} \int_0^R \hat{\omega} (t) J_{\frac{1}{2}} (\lambda t) \, \Intd t \, , \label{solutionForm}
\end{equation}
wherein $\hat{\omega}(t)$, with $t \in [0, R]$, is an unknown function later to be determined.
We will show in the sequel that the equation for the outer problem~\eqref{outerProblem} is indeed satisfied using this form of solution.

First, it can readily be checked that Eq.~\eqref{solutionForm} can further be expressed in the form
\begin{equation}
	\omega(\lambda) = 
	\lambda^{-\frac{3}{2}} \int_0^R \hat{\omega}(t) \, t^{-\frac{3}{2}} 
	\frac{\Intd}{\Intd t} \left( t^\frac{3}{2} J_\frac{3}{2} (\lambda t) \right) \Intd t \, .  \label{solutionFormRewritten}
\end{equation}

By defining
\begin{equation}
	\hat{F} (t) = t^\frac{3}{2} \,
	\frac{\Intd}{\Intd t} \left( t^{-\frac{3}{2}} \hat{\omega} (t) \right)
\end{equation}
and assuming that $t^\frac{3}{2} \, \hat{\omega} (t) \to 0$ as $t \to 0^+$, 
Eq.~\eqref{solutionFormRewritten} can be rewritten upon integration by parts as
\begin{equation}
	\omega(\lambda) = \lambda^{-\frac{3}{2}} 
	\left( \hat{\omega}(R) J_\frac{3}{2} (\lambda R)
	- \int_0^R \hat{F}(t) J_\frac{3}{2} (\lambda t) \, \Intd t \right) .
	\label{solutionFormRewrittenFinal}
\end{equation}

Then, by substituting the modified form of solution given by Eq.~\eqref{solutionFormRewrittenFinal} into Eq.~\eqref{outerProblem}, the integral equation for the outer problem can be expressed as
\begin{equation}
	\mathcal{K}_+ (r, R) \, \hat{\omega} (R)
	- \int_0^R \mathcal{K}_+ (r,t) \hat{F} (t) \, \Intd t = 0 
	\quad (r > R)	\, . \label{outerTransformed}
\end{equation}
In this context, we define the kernel functions
\begin{equation}
	\mathcal{K}_\pm (r,t) = \int_0^\infty \lambda^\frac{1}{2}  J_{1 \pm \frac{1}{2}} (\lambda t) J_1(\lambda r) \, \Intd \lambda \, . \label{KpmDef}
\end{equation}
It turned out that the latter improper (infinite) integral can be evaluated analytically as
\begin{equation}
	\mathcal{K}_+ (r, t) 
	= \left( \frac{2}{\pi t} \right)^\frac{1}{2} \frac{r}{t}
	\frac{\Theta(t-r)}{\left( t^2 - r^2 \right)^{\frac{1}{2}}} \, , 
\end{equation}
with $\Theta(\cdot)$ denoting the Heaviside step function (or the unit step function).
Since $r>R$, it can readily be perceived that the transformed integral equation for the outer problem stated by Eq.~\eqref{outerTransformed} is trivially satisfied.

Thereafter, substituting Eq.~\eqref{solutionForm} into the integral equation for the inner problem given by Eq.~\eqref{innerProblem} yields 
\begin{equation}
	\int_0^R \mathcal{K}_- (r, t) \, \hat{\omega}(t) \, \Intd t 
	 = f(r) \qquad (r<R) \, ,  \label{innerProblemSubs}
\end{equation}
By noting that
\begin{equation}
\mathcal{K}_- (r, t) = \left( \frac{2t}{\pi} \right)^\frac{1}{2} \frac{1}{r}
	\frac{\Theta(r-t)}{\left( r^2-t^2 \right)^\frac{1}{2}} \, , \label{Km}
\end{equation}
Eq.~\eqref{innerProblemSubs} can subsequently be rewritten in a much simplified form as
\begin{equation}
	\int_0^r \frac{t^\frac{1}{2} \hat{\omega} (t) }{\left( r^2-t^2 \right)^\frac{1}{2}} \, \Intd t
	= \left( \frac{\pi}{2} \right)^\frac{1}{2} r f(r) \qquad (r < R) \, .  \label{Abel}
\end{equation}

Equation~\eqref{Abel} is a classical Abel integral equation which constitutes a special form of the linear Volterra equation of the first kind having
a weakly singular kernel\cite{carleman21, smithies58, anderssen80}.
It admits a unique solution if and only if $f(r)$ is a continuously differentiable function~\cite{whittaker96, carleman22, tamarkin30}.
Its solution is formally given in an integral form as  (c.f.\ Appendix~\ref{appendixAbel} for further details)
\begin{equation}
	\hat{\omega}(t) = \left( \frac{2}{\pi t} \right)^\frac{1}{2}
	\frac{\Intd}{\Intd t} \int_0^t \frac{v^2 f(v) \, \Intd v }{\left( t^2 - v^2 \right)^\frac{1}{2}}  \, .
	\label{AbelInter}
\end{equation}
Inserting the expression of $f(r)$ stated by Eq.~\eqref{radialFct} into Eq.~\eqref{AbelInter} and performing the resulting integration yields
\begin{equation}
	\hat{\omega}(t)	= -2 \left( \frac{2t}{\pi} \right)^\frac{1}{2} 
	\frac{ht}{\left( t^2 + h^2 \right)^2} \, . \label{hatomega}
\end{equation}

Evidently, the condition $t^\frac{3}{2} \, \hat{\omega} (t) \to 0$ as $t \to 0^+$ assumed above upon integrating by parts is well satisfied.

Next, by substituting Eq.~\eqref{hatomega} into Eq.~\eqref{solutionForm} upon noting that
\begin{equation}
	J_\frac{1}{2} \left( \lambda t \right) = 
	\left( \frac{2}{\pi} \right)^\frac{1}{2} 
	\left( \lambda t \right)^{-\frac{1}{2}}
	\sin \left( \lambda t \right) \, ,
\end{equation}
the unknown wavenumber-dependent function $\omega(\lambda)$ can be written in a compact integral form as
\begin{equation}
	\omega(\lambda) = -\frac{4h}{\pi \lambda}
	\int_0^R \frac{t\sin (\lambda t) \, \Intd t }{\left( t^2+h^2 \right)^2}  \, .
	\label{omegaLambdaSol}
\end{equation}
The latter integral can be expressed in a general way in terms of hyperbolic functions and trigonometric integrals.
However, choosing the integral form is more convenient for later treatment.
In particular, it follows that $\omega(\lambda) = -e^{-\lambda h}$ as $R \to \infty$.
Finally, by inserting Eq.~\eqref{omegaLambdaSol} into Eq.~\eqref{vIm} and interchanging the order of integration with respect to the variables $\lambda$ and~$t$, the solution for the image azimuthal velocity field follows as
\begin{equation}
	v_\phi^* (r, z) = -\frac{4Kh}{\pi} \int_0^R \frac{t \, \mathcal{Q} (r, z, t) \, \Intd t}{\left( t^2 + h^2 \right)^2} \, , \label{vPhiFinal}
\end{equation}
where we have defined the kernel function
\begin{equation}
	\mathcal{Q} (r, z, t) = \int_0^\infty J_1(\lambda r) \sin(\lambda t) \, e^{-\lambda |z|}  \, \Intd  \lambda \, . \label{Q_def}
\end{equation}

To obtain an analytical expression for $\mathcal{Q}$, we proceed by making use of Euler's formula in complex analysis~\cite{ahlfors1966complex} and write $\sin (\lambda t) = \operatorname{Im} \left\{ e^{i\lambda t} \right\}$, with Im denoting the imaginary part of the argument.
By using the change of variable $u = \lambda r$, Eq.~\eqref{Q_def} can be rewritten in the form
\begin{equation}
	\mathcal{Q} (r, z, t) = \frac{1}{r} \operatorname{Im} \left\{
	\int_0^\infty J_1(u) \, e^{-su} \, \Intd u 	\right\} , \label{Q_rewritten}
\end{equation}
with $s = \left( |z|-it \right) / r$.
By invoking the Laplace transform~\cite{widder15} of $J_1(u)$ given as $1 - s\left( 1 + s^2 \right)^{-\frac{1}{2}}$, Eq.~\eqref{Q_rewritten} can then be evaluated as
\begin{equation}
	\mathcal{Q} (r, z, t) = -\frac{1}{r} \operatorname{Im} 
	\left\{ \frac{|z| - it}{\left( r^2 + \left( |z|-it \right)^2 \right)^\frac{1}{2}} \right\} .
\end{equation}
The latter can further be cast in the final form as
\begin{equation}
	\mathcal{Q} (r, z, t) =
	\frac{2^\frac{1}{2}}{2rU} 
	\left( t \left( U+V \right)^\frac{1}{2} - |z| \left( U-V \right)^\frac{1}{2} \right),
\end{equation}
where we have defined
\begin{equation}
	U = \left( \left( \rho^2+t^2 \right)^2 - 4 r^2 t^2 \right)^\frac{1}{2} \, , \qquad
	V = \rho^2 - t^2 \, , 
\end{equation}
where $\rho^2 = r^2 + z^2$.
It can be shown that $\mathcal{Q}$ is always well defined except when $U=0$, for which $z=0$ and $t=r$. 
In this case, $\mathcal{Q} (r,z,t) \sim \left( r-t \right)^{-\frac{1}{2}} \Theta \left( r-t \right)$.
We note that $Q(r,z,t) \to 0$ as $r \to 0$.

An analytical integration of Eq.~\eqref{vPhiFinal} is delicate, if not downright impossible.
Therefore, recourse to numerical procedures is necessary.
To this end, we approximate the integral by a standard middle Riemann sum using the partition $t_1, \dots, t_{N}$, where $t_i = \left( i - 1/2 \right)\delta$, with $i = 1, \dots, N$ and $\delta = R/N$.
Here, $N$ denotes the number of discretization points.
Throughout this work, we consistently set $N=10\,000$.

\begin{figure*}
	\includegraphics[scale=0.285]{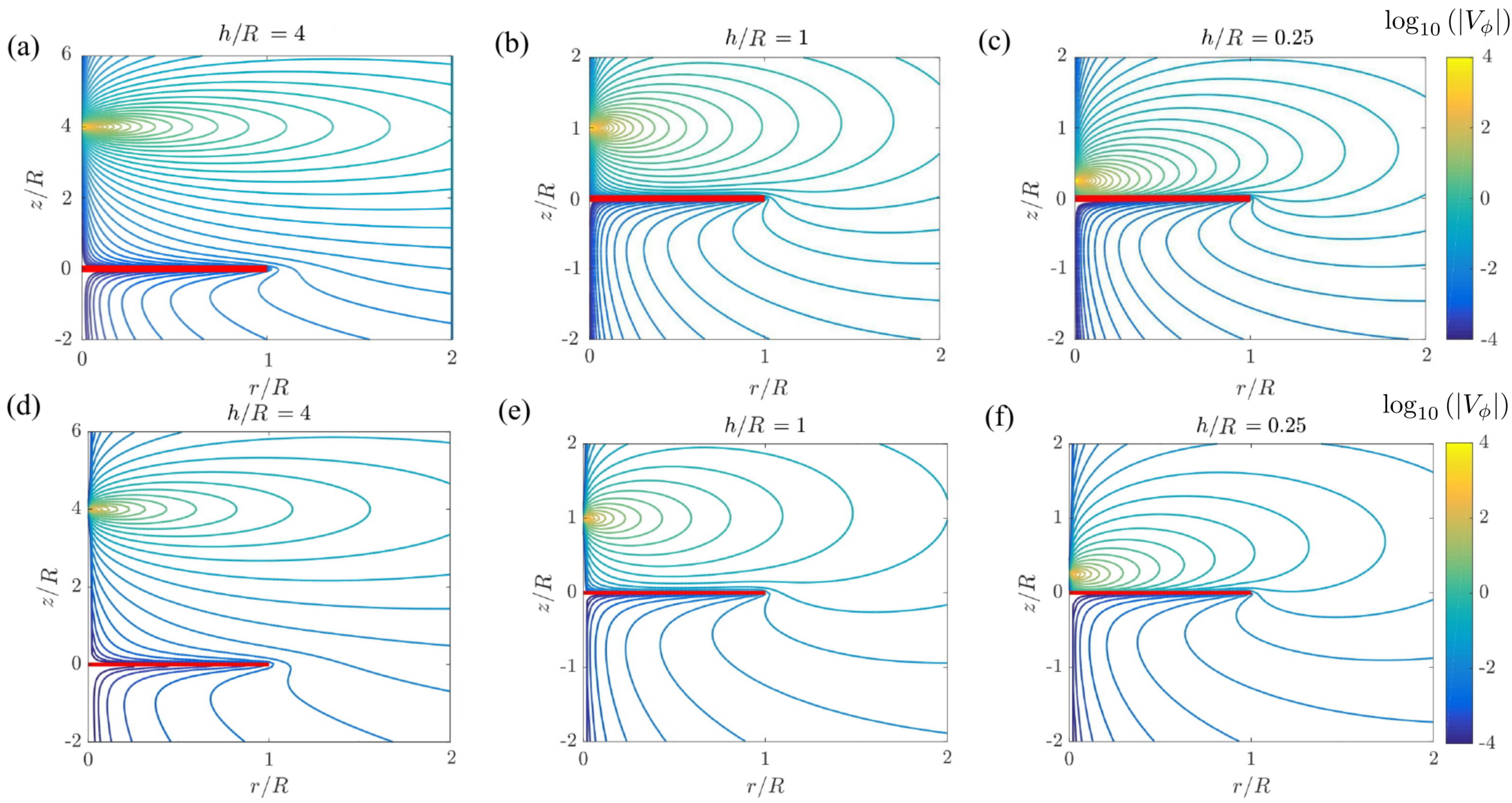}
	\caption{(Color online) Contour plot of the amplitude of the scaled azimuthal flow velocity as obtained theoretically [first row, $(a)$, $(b)$, and $(c)$] and using finite-element simulations [second row, $(d)$, $(e)$, and $(f)$].
	The flows are induced by a rotlet singularity positioned at $h/R = 4$ [$(a)$ and~$(d)$], $h/R = 1$ [$(b)$ and~$(e)$] and $h/R = 0.25$ [$(c)$ and~$(f)$] on the axis of a no-slip disk of radius $R$ (red).
	The scaled azimuthal velocity is defined as $V_\phi = v_\phi / \left( L/\left( 8\pi\eta R^2 \right) \right)$.
	Results are presented on a decimal logarithmic scale.
	}
	\label{ContourPlotSingleDisk}
\end{figure*}

In Fig.~\ref{ContourPlotSingleDisk}, we show contour plots of the scaled azimuthal flow velocity induced by a rotlet singularity positioned at different
distances along the axis of the disk.
The analytical predictions [$(a)$, $(b)$, and~$(c)$] are in good agreement with the results from finite-element simulations [$(d)$, $(e)$, and~$(f)$; see Appendix~\ref{appendixRichter} for technical details regarding the simulation method].

The magnitude of the scaled velocity field is shown on a logarithmic scale to emphasize the difference between the different regions.
Similarly, as in the bulk, the flow velocity is decaying faster along
the $z$ direction (or axis of rotation) compared to the radial direction.
However, near the disk, the azimuthal flow velocity becomes asymmetric
with respect to the rotlet position, and for $z<0$ the total flow field
almost vanishes completely.
Last, we remark that the overall magnitude of the flow field reduces as
the rotlet gets closer to the disk.

\subsection{Exact solution for $R \to \infty$}

The solution for a rotlet oriented normal to a hard wall can be obtained using the image system technique noted by Blake~\cite{blake71}.
For completeness, we here derive this solution in a different way using our formalism. 
For an infinitely large disk located at $z=0$, the integral equations~\eqref{innerProblem} for the inner domain hold for all positive~$r$, and thus we can directly apply a Hankel transform~\cite{piessens2000hankel} on both sides of this equation. 
Here, we make use of the orthogonality property of Bessel functions~\cite{abramowitz72}
\begin{equation}
	\int_{0}^{\infty} r J_{\nu}(\lambda r)  J_{\nu}(\lambda' r) \, \Intd r 
	= \lambda^{-1} \delta (\lambda - \lambda') \, ,
\end{equation}
and obtain the solution for
\begin{equation}
	\omega(\lambda) = \int_0^\infty r f(r) J_1(\lambda r) \, \Intd r  = -e^{-\lambda h} \, . 
\end{equation}

Inserting the latter result into Eq.~\eqref{vIm} yields
\begin{equation}
	{v_\phi^*}_\text{Blake} (r, z) = -\frac{K r}{\left( r^2 + \left ( |z| + h \right)^2 \right)^{\frac{3}{2}}} \, .
	\label{Blake}
\end{equation}
The image solution for the azimuthal component in the limit $R\to\infty$ can alternatively be calculated analytically from Eq.~\eqref{vPhiFinal}.
Consequently, the total flow field vanishes underneath the disk, i.e. for $z<0$.

\begin{figure}
	\includegraphics[scale=1]{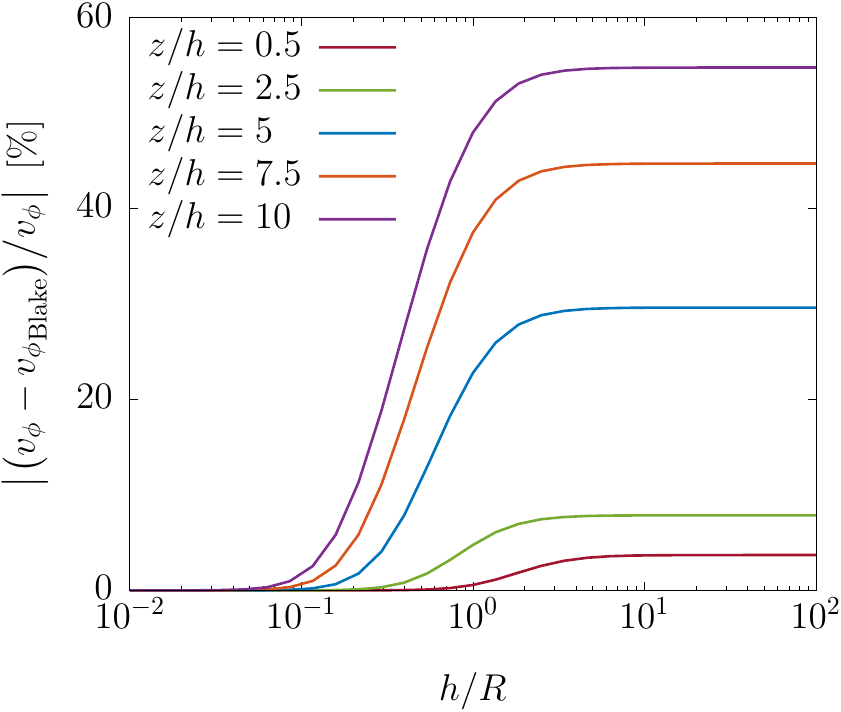}
	\caption{
	(Color online) Percentage relative error in the flow field as obtained using Blake's solution for an infinitely extended disk in comparison to the exact solution for a disk of finite size, as a function of increasing vertical distance $h/R$ of the rotlet from the disk. 
	The flow field is evaluated at five different values of~$z/h$ while keeping $r/R = 0.2$.}
	\label{Error}
\end{figure}

Figure~\ref{Error} shows the variation of the percentage relative error in the flow velocity field as obtained using Blake's solution given by Eq.~\eqref{Blake} in comparison to the exact solution derived in the present work for a finite-sized disk given in an integral form by Eq.~\eqref{vPhiFinal}.
Here, the flow field is evaluated at five different axial distances~$z/h$, while keeping the radial position $r/R = 0.2$.
We observe that the error is vanishingly small for $h \ll R$ and increases monotonically as the ratio $h/R$ gets larger.
In addition, the error amounts to small values in the fluid domain close to the singularity position in which the flow velocity is primarily determined by the infinite-space rotlet.
Upon increasing $z/h$, the maximum percentage relative error (MPRE) increases and it was found to be as high as approximately 55\% for $z/h = 10$ and~$h/R = 10^2$.
We have systematically checked that the MPRE is, in general, less sensitive to variations in the radial position.

\subsection{Hydrodynamic rotational mobility}

Having derived the solution of the flow problem for a point-torque singularity acting near a finite-sized disk, we next investigate how the presence of the nearby disk affects the rotational mobility.
For this purpose, we think of the rotlet generated by a small colloidal particle of radius~$a$.
By restricting ourselves to the situation in which $a \ll h$, the leading-order correction to the particle rotational mobility can be obtained by evaluating the image angular velocity of the fluid, $\tfrac{1}{2} \, \boldsymbol{\nabla} \times \vect{v}^*$, at the singularity position~\cite{swan07}.
In a scaled form, it can be presented as
\begin{equation}
	\frac{\Delta \mu}{\mu_0} = \frac{a^3}{K} \lim\limits_{(r,z) \to (0, h)} 
	\frac{1}{2r} \frac{\partial }{\partial r} \left( r v_\phi^* \right) \, , 
	\label{mobiCorrDef}
\end{equation}
wherein $\mu_0 = \left( 8\pi\eta a^3 \right)^{-1}$ is the bulk rotational mobility, i.e., in the absence of the disk.

Following the notation employed in our previous considerations~\cite{daddi19jpsj, daddi2020asymmetric, daddi2020dynamics}, we define the positive dimensionless number~$k_1$ to be the scaled correction factor to the mobility near a no-slip disk as
\begin{equation}
	k_1 = -\left. \frac{\Delta \mu}{\mu_0} \middle/ \left( \frac{a}{h} \right)^3 \right. , 
\end{equation}
and substituting Eq.~\eqref{vIm} expressing the image azimuthal velocity into Eq.~\eqref{mobiCorrDef}, we obtain
\begin{equation}
	k_1 = -\frac{h^3}{2} \int_0^\infty \lambda^2 \omega(\lambda) e^{-\lambda h} \, \Intd \lambda \, . \label{kExp}
\end{equation}

Next, by inserting the expression of~$\omega(\lambda)$ stated by Eq.~\eqref{omegaLambdaSol} into Eq.~\eqref{kExp} and using the changes of variables $u = \lambda h$ and $v = t/R$, the scaled correction factor can be presented as an integral over the interval $[0,1]$ as
\begin{equation}
	k_1 (\xi) = \frac{2}{\pi} \, \xi^2 
	\int_0^1 \frac{v \, G(v, \xi) \, \Intd v}{\left( v^2 + \xi^2 \right)^2}  \, .
	\label{kAlmostThere}
\end{equation}
with the dimensionless number $\xi = h/R$.
Here,
\begin{equation}
	G(v, \xi) = \int_0^\infty u \sin \left( \frac{uv}{\xi} \right) e^{-u} \, \Intd u
	= \frac{2v\xi^3}{\left( v^2 + \xi^2 \right)^2} \, .
\end{equation}

Finally, evaluating the definite integral in Eq.~\eqref{kAlmostThere} yields the expression of the scaled correction factor as
\begin{equation}
	k_1 (\xi) = \frac{1}{8} - \frac{1}{4\pi} 
	\left( \arctan \xi + \frac{\xi \left(\xi^2-3\right) \left( 1+3\xi^2 \right) } {3 \left( 1+\xi^2 \right)^3} \right) . \label{k1Final}
\end{equation}
In particular, for $\xi \ll 1$ (or $h \ll R$), we obtain
\begin{equation}
	k_1 (\xi) = \frac{1}{8} - \frac{4}{5\pi} \, \xi^5 + \mathcal{O} \left( \xi^7 \right) \, .
\end{equation}
Notably, the familiar correction factor $k_1 = 1/8$ near an infinitely extended hard wall is recovered in the limit $\xi \to 0$.
For $\xi \gg 1$, we get
\begin{equation}
	k_1 (\xi) = \frac{4}{3\pi} \, \xi^{-3} + \mathcal{O} \left( \xi^{-5} \right) \, .
\end{equation}
Interestingly, the correction factor takes a particularly simple expression when $\xi = \sqrt{3}$, for which $k_1 = 1/24$.

\begin{figure}
	\centering
	\includegraphics[scale=1]{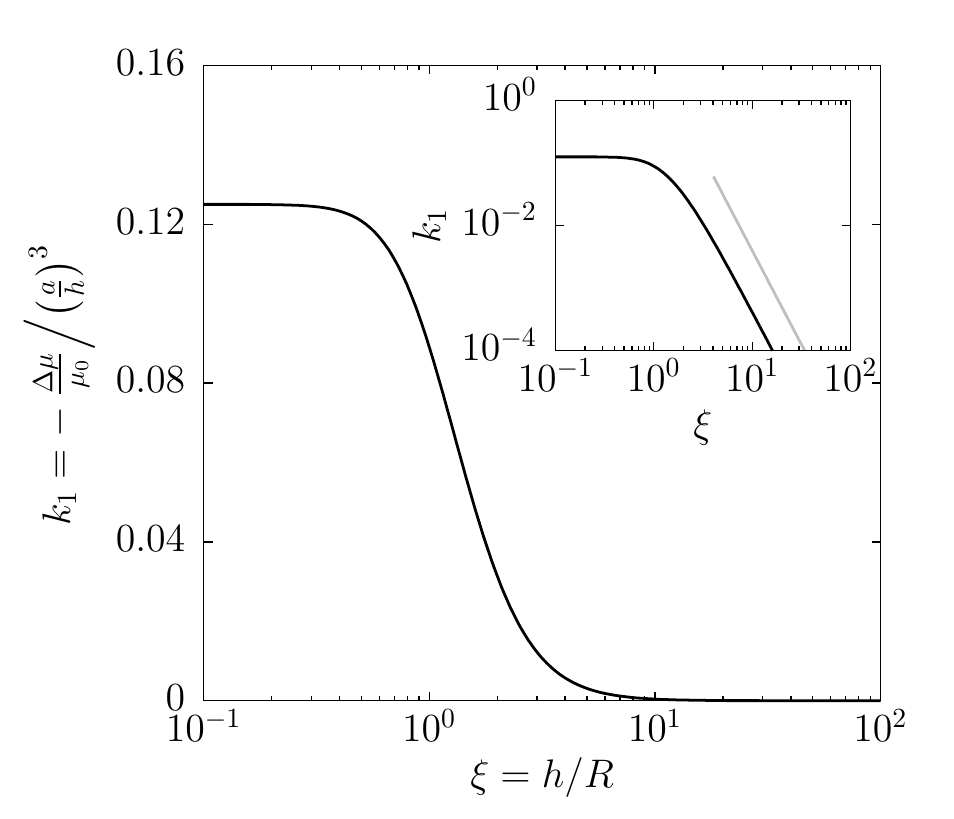}
	\caption{Scaled correction factor~$k_1$ to the rotational hydrodynamic mobility near a rigid no-slip disk, Eq.~\eqref{k1Final}, versus the dimensionless number~$\xi = h/R$ plotted on a semilogarithmic scale.
	The same curve is shown in the inset on a log-log scale, where the scaling law~$\xi^{-3}$ is displayed in the range $\xi \gg 1$ (gray).
	}
	\label{MobiSingle}
\end{figure}

In Fig.~\ref{MobiSingle}, we present the variation of the scaled correction factor given by Eq.~\eqref{k1Final} as a function of the dimensionless number~$\xi = h/R$.
We observe that the scaled correction factor is a monotonically decaying function of~$\xi$.
On a semilogarithmic scale (main plot), the curve exhibits an inverse logistic-like (sigmoid) evolution between two plateau values.
The scaled correction factor undergoes a cubic decay with~$\xi$ while vanishing in the limit $\xi \to \infty$.

Recapitulating, we have presented a dual integral equation approach to determine the solution of the hydrodynamic equations for a point-torque singularity acting near a rigid disk.
In the following, we will employ a similar technique to obtain the corresponding solution of the flow problem in the presence of two coaxially positioned rigid disks.

\section{Solution for two coaxially positioned disks}
\label{sec:twoDisks}

\begin{figure}
	\centering
	\includegraphics[scale=1.2]{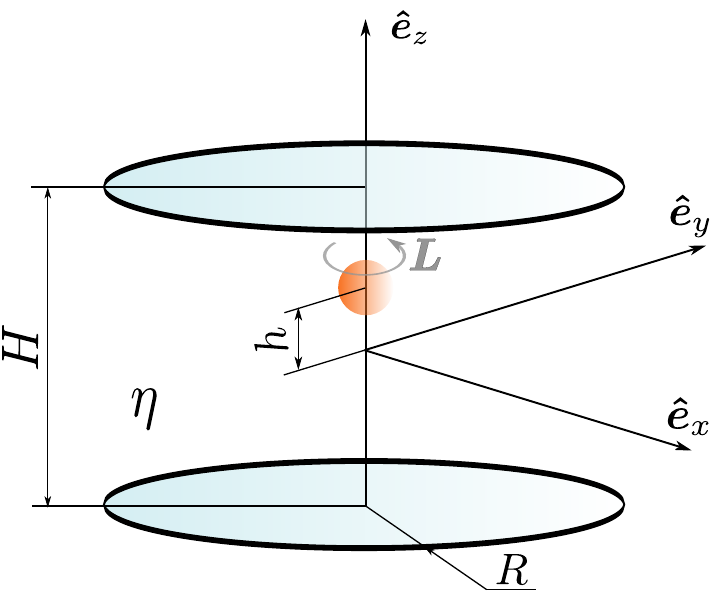}
	\caption{(Color online) Slow rotational motion of a point-like particle confined between two coaxially positioned rigid no-slip disks of identical radius~$R$.
	The confining disks are located within the planes $z = \pm H/2$ with~$H$ denoting the separation distance between the parallel disks.
	The particle is located at a distance~$h$ from the origin of the system of coordinates on the axis of the disks.
	}
	\label{IllusChannel}
\end{figure}

\subsection{Problem formulation}

We now assume that two parallel coaxially positioned rigid disks are located within the planes at $z = -H/2$ and $z = H/2$, where~$H$ represents the distance separating the two disks.
The $z$~axis passes through the centers of the coaxially positioned disks.
We suppose that the rotlet is acting between the disks at position $\R_0 = h \vect{e}_z$, where $-H/2 < h < H/2$; see Fig.~\ref{IllusChannel} for a graphical illustration of the setup.

To find a solution to the flow problem, we partition the fluid medium into three distinct parts.
We label by the superscript 1 the flow velocity field in the fluid domain beneath the plane $z = -H/2$ containing the lower disk, subscript 2 the fluid region delimited by the planes at $z = -H/2$ and $z = H/2$, and we designate by the subscript 3 the fluid domain above the plane $z = H/2$ containing the top disk.
In the remainder of this article, we choose, for convenience, to scale all lengths by the gap width~$H.$

We now express the solution for the azimuthal velocity field in each region of the fluid domain as
\begin{subequations} \label{velocityChannel}
	\begin{align}
		{v_\phi^*}_1 &= K \int_0^\infty 
		A(\lambda)  e^{\lambda z} J_1(\lambda r) \, \Intd \lambda \, , \\
		{v_\phi^*}_2 &= K \int_0^\infty 
		\left( B(\lambda)  e^{-\lambda z} + C(\lambda) e^{\lambda z} \right) J_1(\lambda r) \, \Intd \lambda \, , \\
		{v_\phi^*}_3 &= K \int_0^\infty 
		D(\lambda)  e^{-\lambda z} J_1(\lambda r) \, \Intd \lambda \, ,
	\end{align}
\end{subequations}
where $A(\lambda)$, $B(\lambda)$, $C(\lambda)$, and $D(\lambda)$ are unknown functions to be determined from the underlying boundary conditions.
It can readily be checked that the regularity condition of a finite velocity field is inherently satisfied in the whole domain.

\subsection{Boundary conditions and dual integral equations}

Requiring the natural continuity of the velocity field at the planes $z = \pm 1/2$ yields the expressions of the functions associated with the intermediate fluid domain in terms of those related to the lower and upper domains.
Specifically,
\begin{subequations} \label{BundC}
	\begin{align}
		B(\lambda) &= \frac{1}{2} \left( A(\lambda) - D(\lambda) e^{-\lambda} \right) \operatorname{csch} (\lambda) \, , \\
		C(\lambda) &= \frac{1}{2} \left( D(\lambda) - A(\lambda) e^{-\lambda} \right) \operatorname{csch} (\lambda) \, ,
	\end{align}
\end{subequations}
with csch denoting the hyperbolic cosecant function, defined as $\operatorname{csch} (\lambda) = 1/\sinh \lambda = 2 / \left( e^\lambda - e^{-\lambda} \right)$.

By imposing the no-slip boundary condition at the surfaces of the two disks, we obtain the equations for the inner problem for $r<R$ as
\begin{subequations} \label{innerEqsChannel}
	\begin{align}
		\int_0^\infty A(\lambda) e^{-\frac{\lambda}{2}} J_1(\lambda r) \, \Intd \lambda &= \psi_+(r) \, , \\
		\int_0^\infty D(\lambda) e^{-\frac{\lambda}{2}} J_1(\lambda r) \, \Intd \lambda &= \psi_-(r) \, , 
	\end{align}
\end{subequations}
where we have defined the radially symmetric functions
\begin{equation}
	\psi_\pm (r) = -\frac{r}{\left( r^2 + \left( h \pm \frac{1}{2} \right)^2 \right)^\frac{3}{2}} \, .
\end{equation}

In addition, the continuity of the azimuthal stress vector outside the regions containing the disk yields the equations for the outer problem for $r > R$.
Specifically,
\begin{subequations}\label{outerEqsChannel}
	\begin{align}
		\int_0^\infty \lambda \left( A (\lambda) e^\frac{\lambda}{2} - D (\lambda) e^{-\frac{\lambda}{2}} \right) \operatorname{csch} (\lambda) J_1 (\lambda r) \, \Intd \lambda &= 0 , \\
		\int_0^\infty \lambda \left( A (\lambda) e^{-\frac{\lambda}{2}} - D (\lambda) e^\frac{\lambda}{2} \right) \operatorname{csch} (\lambda) J_1 (\lambda r) \, \Intd \lambda &= 0 .
	\end{align}
\end{subequations}

Equations~\eqref{innerEqsChannel} and \eqref{outerEqsChannel} constitute a system of dual integral equations for the unknown functions $A(\lambda)$ and~$D(\lambda)$.
For its solution, we employ the standard solution approach outlined by Sneddon~\cite{sneddon60} and Copson~\cite{copson61} and set
\begin{subequations} \label{f1f2Defs}
	\begin{align}
		\frac{1}{2} \left( A (\lambda) e^\frac{\lambda}{2} - D (\lambda) e^{-\frac{\lambda}{2}} \right) \operatorname{csch} (\lambda) &= 
		\lambda^\frac{1}{2} f_1(\lambda) \, , \\
		\frac{1}{2} \left( A (\lambda) e^{-\frac{\lambda}{2}} - D (\lambda) e^\frac{\lambda}{2} \right) \operatorname{csch} (\lambda) &= 
		\lambda^\frac{1}{2} f_2(\lambda) \, , 
	\end{align}
\end{subequations}
where 
\begin{equation}
	f_i(\lambda) = \int_0^R \hat{f}_i(t) J_\frac{1}{2} (\lambda t) \, \Intd t \, ,
\end{equation}
with $\hat{f}_i (t), i \in \{ 1, 2\}$, are two unknown functions defined on the interval $[0, R]$ to be subsequently determined. 
In this way, the equations for the outer problem are automatically satisfied following the same reasoning in Sec.~\ref{sec:singleDisk}.
Solving Eqs.~\eqref{f1f2Defs} for $A(\lambda)$ and~$D(\lambda)$ yields
\begin{subequations} \label{AundD}
	\begin{align}
		A(\lambda) = \lambda^\frac{1}{2}
		\left( f_1(\lambda) e^{\frac{\lambda}{2}} - f_2(\lambda) e^{-\frac{\lambda}{2}}  \right) , \\
		D(\lambda) = \lambda^\frac{1}{2}
		\left( f_1(\lambda) e^{-\frac{\lambda}{2}} - f_2(\lambda) e^{\frac{\lambda}{2}}  \right) .
	\end{align}
\end{subequations}

Upon substitution of Eqs.~\eqref{AundD} into Eqs.~\eqref{innerEqsChannel}, the inner problem can be expressed in the form
\begin{subequations} \label{SystDualIntEqns}
	\begin{align}
		\int_0^R \left( \mathcal{K}_-(r,t) \hat{f}_1(t)
		- \mathcal{S} (r, t) \hat{f}_2(t)	 \right) \Intd t &= \psi_+ (r) \, , \\
		\int_0^R \left( \mathcal{S}(r,t) \hat{f}_1(t)
		- \mathcal{K}_- (r, t) \hat{f}_2(t)	 \right) \Intd t &= \psi_- (r) \, .
	\end{align}
\end{subequations}
Here, we have defined the kernel function
\begin{equation}
	\mathcal{S} (r, t) =
	\left( \frac{2}{\pi t} \right)^\frac{1}{2} \mathcal{Q} (r, 1, t) \, ,
	\label{S}
\end{equation}
where $\mathcal{Q}$ has been defined earlier by Eq.~\eqref{Q_rewritten}.
We note that the expression of the kernel function $\mathcal{K}_{-}$ has previously been given by Eq.~\eqref{Km} and can further be expressed in term of $\mathcal{Q}$ as
\begin{equation}
	\mathcal{K}_- (r, t) =
		\left( \frac{2}{\pi t} \right)^\frac{1}{2} \mathcal{Q} (r, 0, t) \, .
\end{equation}

Equations~\eqref{SystDualIntEqns} represent a system of Fredholm integral equations of the first kind~\cite{tricomi85}.
Due to the complicated expressions of their kernel functions, exact analytical expressions for the unknown functions $\hat{f}_1(t)$ and~$\hat{f}_2(t)$ are far from trivial.
Following a computational approach, we partition the integration intervals~$[0, R]$ into~$N$ subintervals, approximating the integrals by the standard  middle Riemann sum.
We then evaluate the two resulting equations at~$N$ discrete values of~$r$ that are distributed uniformly over the interval $[0,R]$.
Inverting the resulting linear system of~$2N$ independent equations, we obtain accurate values of $\hat{f}_1(t)$ and~$\hat{f}_2(t)$ at each discretization point.

Inserting the expressions of $A(\lambda)$ and~$D(\lambda)$ stated by Eqs.~\eqref{AundD} into Eqs.~\eqref{velocityChannel}, the image solution for the azimuthal flow field everywhere in the fluid domain can be cast in the final compact form
\begin{equation}
	v_\phi^*(r, z) = K \int_0^R \left( \frac{2}{\pi t} \right)^\frac{1}{2} \hat{g} (r,z,t) \, \Intd t \, , \label{vPhiStarChannel}
\end{equation}
wherein
\begin{equation}
	\hat{g} (r,z,t) = \mathcal{Q} \left( r, z + \frac{1}{2}, t \right) \hat{f}_1(t)
	- \mathcal{Q} \left( r, z - \frac{1}{2}, t \right) \hat{f}_2(t) \, . \notag
\end{equation}

Notably, the system of Fredholm integral equations stated by Eqs.~\eqref{SystDualIntEqns} is recovered when enforcing the no-slip condition at $z = \pm 1/2$.

Likewise, the definite integral given by Eq.~\eqref{vPhiStarChannel} can be discretized via the middle Riemann sum to yield an approximate solution for the image velocity field at any point $(r,z)$ in the entire fluid domain.

\begin{figure*}
	\includegraphics[scale=0.27]{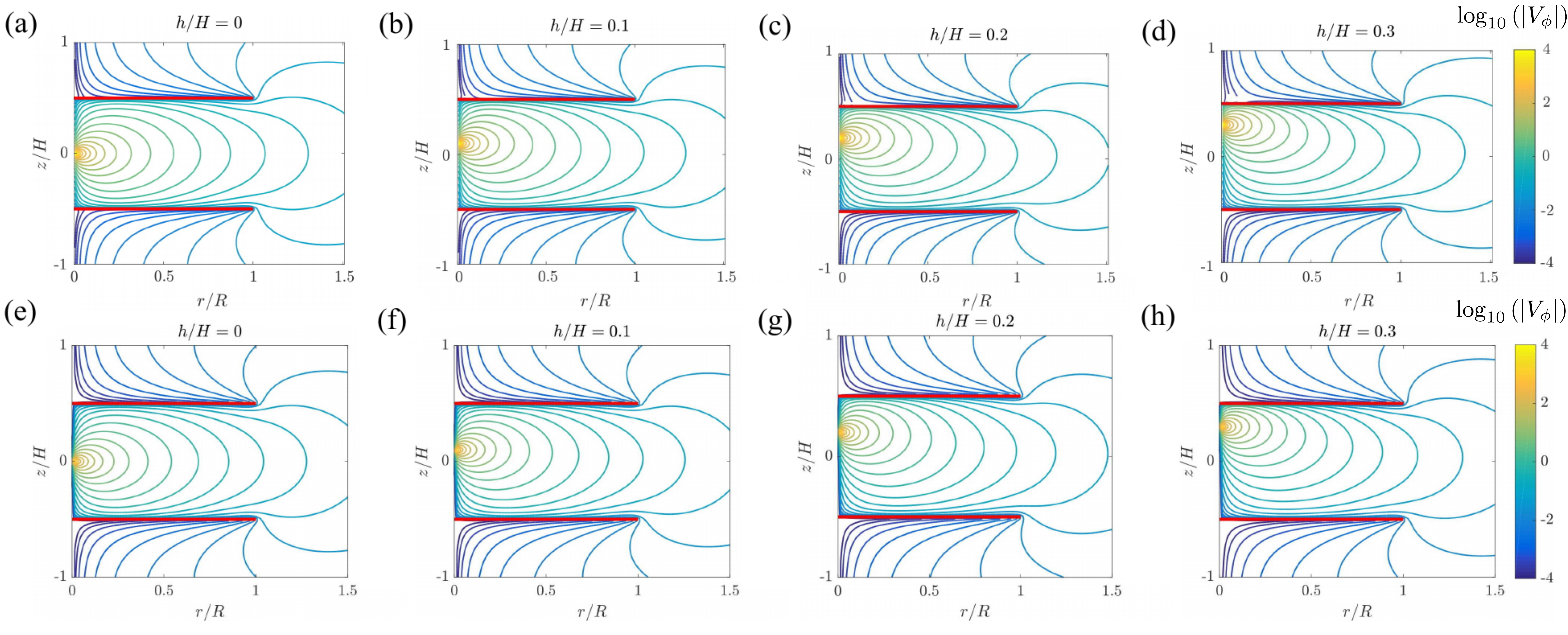}
	\caption{(Color online) Contour plots of the amplitude of the scaled azimuthal velocity as obtained semi-analytically [first row, $(a)$ through~$(d)$] and by means of finite-element simulations [second row, $(e)$ through~$(h)$].
	Results are shown for a rotlet acting at $h/H = 0$ [$(a)$ and~$(e)$], $h/H = 0.1$ [$(b)$ and~$(f)$], $h/H = 0.2$ [$(c)$ and~$(g)$] and $h/H = 0.3$ [$(d)$ and~$(h)$] on the axis of two coaxially positioned no-slip disks of radius $R=H$ (red).
	Here, the scaled azimuthal velocity is defined as $V_\phi = v_\phi / \left( L/\left( 8\pi\eta R^2 \right) \right)$ and results are presented on a decimal logarithmic scale.
	}
	\label{ContourPlotChannel}
\end{figure*}

Figure~\ref{ContourPlotChannel} shows contour plots of the amplitude of the scaled
azimuthal flow velocity for different positions on the axis of two
coaxial disks, as obtained analytically and by means of finite-element simulations. 
Again, the flow velocity is the highest in the near
vicinity of the rotlet and decays faster along the $z$ direction compared
to the radial direction.
Moreover, the magnitude is substantially smaller above and below the disks.
As the rotlet approaches one side of the confining disks, the
overall magnitude of the flow field becomes reduced and the
structure becomes more asymmetric.
Good agreement is obtained between the semi-analytical theory and the finite-element simulations.

\subsection{Solution for $R \to \infty$}

For completeness, we additionally address by our approach the solution for the flow field in a gap bounded by two infinitely extended planar walls located at $z = \pm 1/2$. To this end, a Hankel transform is applied on both sides of Eqs.~\eqref{innerEqsChannel}.
We obtain
\begin{subequations} \label{AundD_Rinf}
\begin{align}
	A(\lambda)  &= \lambda e^{\frac{\lambda}{2}} \bar{\psi}^{+}(\lambda) =
	- \lambda e^{-\lambda h} , \\
	D(\lambda)  &= \lambda e^{\frac{\lambda}{2}} \bar{\psi}^{-}(\lambda) =
	-\lambda e^{\lambda h} ,
\end{align}
\end{subequations}
with
\begin{equation}
	\bar{\psi}^{\pm}(\lambda) = \int_{0}^{\infty} r  \psi^{\pm}(r) J_{1}(\lambda r) \, \Intd r  = - e^{-\lambda (\frac{1}{2} \pm h) },
\end{equation}
for $|h|<1/2$.
Then, it follows from Eqs.~\eqref{BundC} that the wavenumber-dependent function associated with the fluid domain bounded by the planes $z = \pm 1/2$ are given by
\begin{subequations} \label{BundD_Rinf}
	\begin{align}
		B(\lambda) &= -\frac{\lambda}{2} 
		\left( e^{-\lambda h} - e^{-\lambda \left( 1-h \right)} \right)
		\operatorname{csch} (\lambda) \, , \\
		C(\lambda) &= -\frac{\lambda}{2} 
		\left( e^{\lambda h} - e^{-\lambda \left( 1+h \right)} \right)
		\operatorname{csch} (\lambda) \, .
	\end{align}
\end{subequations}

\begin{figure}
	\includegraphics[scale=0.25]{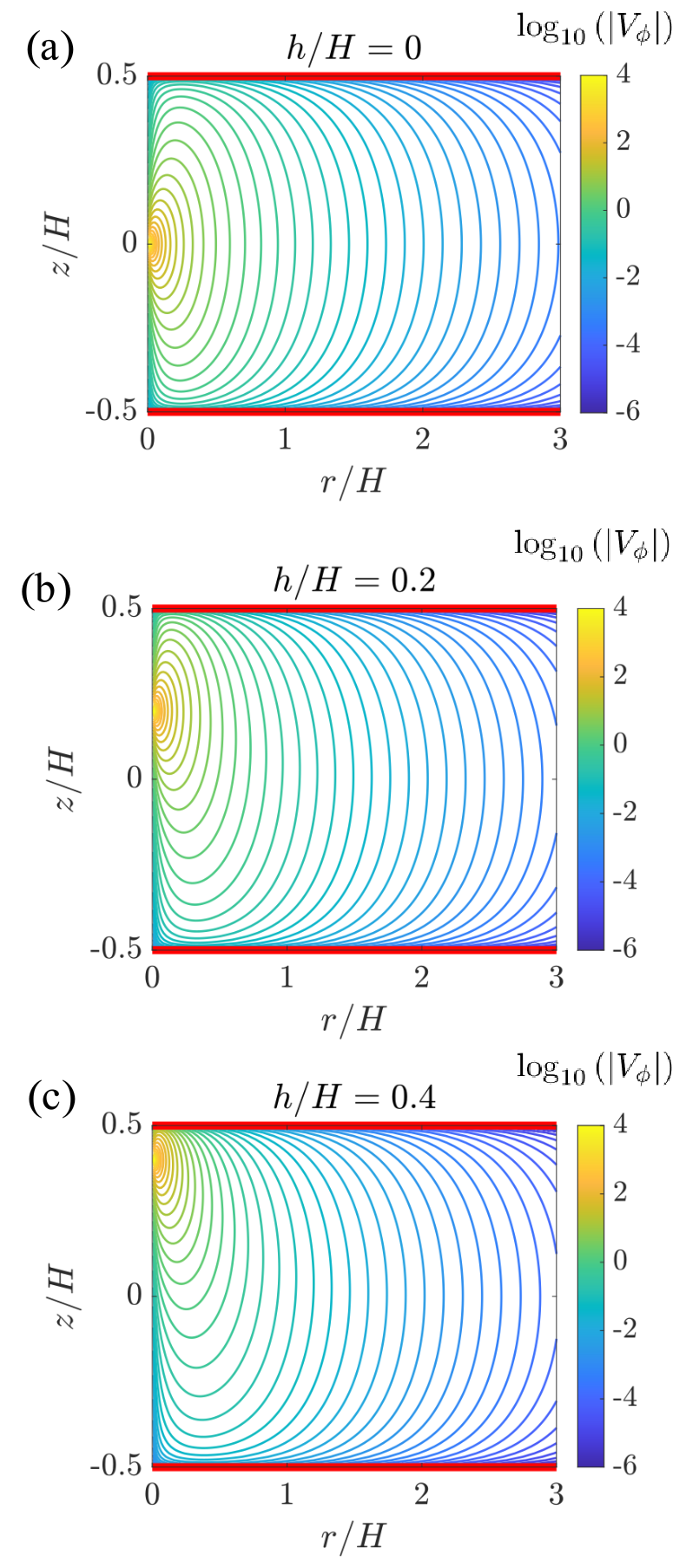}
	\caption{
	(Color online) Contour plots of the amplitude of the scaled azimuthal velocity induced by a rotlet singularity acting between two infinitely extended no-slip walls.
	Again, the scaled azimuthal velocity is defined as $V_\phi = v_\phi / \left( L/\left( 8\pi\eta R^2 \right) \right)$.	
}
	\label{fig7}
\end{figure}

Finally, the corresponding solution of the image flow field can be
obtained by inserting Eqs.~\eqref{AundD_Rinf} and \eqref{BundD_Rinf} into Eqs. \eqref{velocityChannel}.
As expected, the total velocity in the lower and upper regions vanishes in the
limit $R\to\infty$.
Figure~\ref{fig7} illustrates exemplary contour plots of the azimuthal flow field induced by a rotlet acting between two infinitely extended no-slip walls for three different positions of the singularity.

\subsection{Hydrodynamic rotational mobility}

The scaled correction to the particle rotational mobility in the point-particle approximation can be obtained from the image velocity field via Eq.~\eqref{mobiCorrDef}.
Between two coaxially positioned disks, we alternatively choose to define the scaled correction factor as
\begin{equation} 
	k_2 =  - \left. \frac{\Delta \mu}{\mu_0} \middle/ a^3 \right. \, .
\end{equation}
Then, it follows from Eq.~\eqref{vPhiStarChannel} that the scaled correction to leading order can conveniently be expressed as
\begin{equation}
	k_2 = \frac{1}{2} \left( \frac{2}{\pi} \right)^\frac{1}{2} \int_0^R 
	t^{-\frac{1}{2}} \big( \chi_-(t) \hat{f}_2(t) - \chi_+(t) \hat{f}_1(t) \big) \Intd t \, , \label{k2}
\end{equation}
where we have defined
\begin{equation}
	\chi_\pm (t) = \operatorname{Im} \left\{ \left( \frac{1}{2} \pm h - it \right)^{-2} \right\} .
\end{equation}
Again, an approximate evaluation of the definite integral given by Eq.~\eqref{k2} can be performed by numerical discretization via the standard middle Riemann sum.

In the limit of an infinitely extended channel $R\to\infty$, we obtain
\begin{equation}
	k_2 = \frac{1}{2} \int_0^\infty \lambda^2 \operatorname{csch} (\lambda)
	\left( \cosh \left( 2\lambda h \right) - e^{-\lambda} \right) \Intd \lambda \, .
\end{equation}
Using computer algebra systems such as Mathematica~\cite{wolfram1991mathematica}, the latter can further be expressed as
\begin{equation}
	k_2 = \frac{1}{8} 
	\left( \zeta \left( 3, \frac{1}{2} + h \right) 
	+ \zeta \left( 3, \frac{1}{2} - h \right) - 2\zeta(3) \right) ,
	\label{k2inf}
\end{equation}
wherein
\begin{equation}
	\zeta \left( s,t \right) = \sum_{n > m \ge 1} n^{-s} m^{-t}
\end{equation}
denotes the double zeta functions with $s>1$ and $t \ge 0$.
Moreover, $\zeta(s)$ is the Riemann zeta function with $s>1$ defined as $\zeta(s) = \sum_{n \ge 1} n^{-s}$.
In particular, $\zeta(3)$ is an irrational number known as Ap\'{e}ry's constant~\cite{van1979proof}.
We quote the famous identity derived by Euler $\zeta(3) = \zeta(2,1)$.
In the mid-plane of the channel, the correction factor reaches its minimum value $k_2(h=0) = \frac{3}{2} \, \zeta(3) \approx 1.8031$.
Performing a Taylor expansion near the upper wall around $h = 1/2$, we obtain
\begin{equation}
	k_2 = \frac{1}{8} \,  \epsilon^{-3}
	+ \frac{3}{2} \, \zeta(5) \epsilon^2
	+ \mathcal{O} \left( \epsilon^4 \right) ,
\end{equation}
where $\epsilon = \frac{1}{2} - h$.

\begin{figure}
	\centering
	\includegraphics[scale=1]{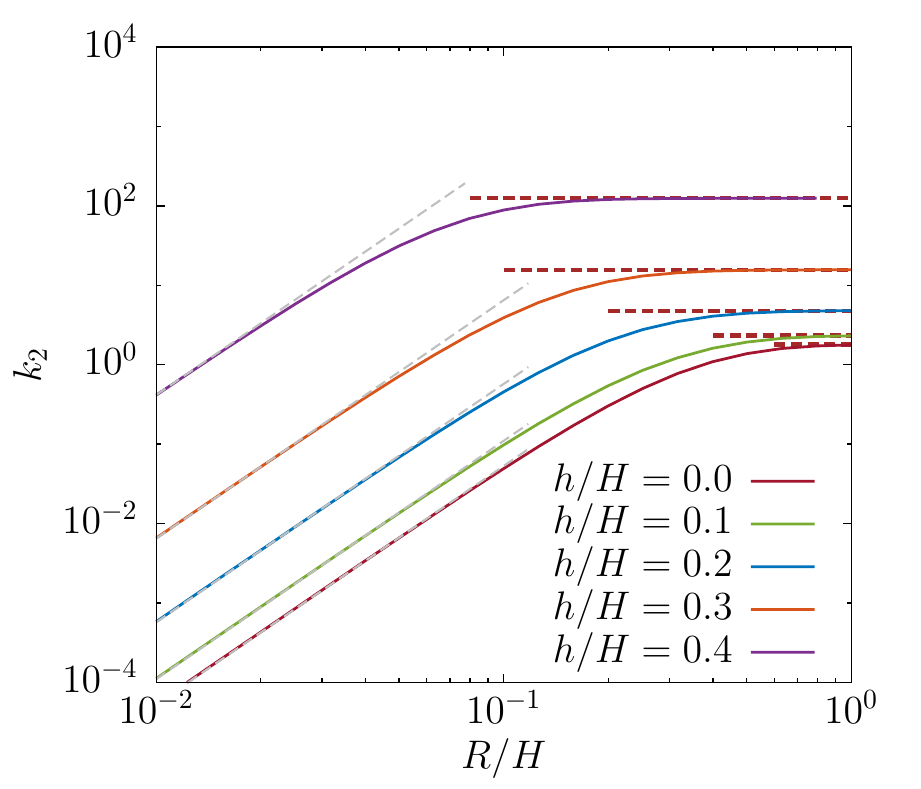}
	\caption{(Color online) Scaled correction factor stated by Eq.~\eqref{k2inf} versus $R/H$ for various singularity positions along the axis of two coaxially positioned rigid no-slip disks.
	Horizontal dashed lines correspond to the scaled correction factors inside an infinitely extended channel.
	The lines shown in gray displays the scaling law $k_2 \sim \left( R/H \right)^3$ in the range of small values of~$R \ll  H$ (c.f.\ Appendix~\ref{appendixMobi} for the derivation of this scaling relation.)
	}
	\label{FigMobiChannel}
\end{figure}

In Fig.~\ref{FigMobiChannel}, we present on a log-log scale the variation of the scaled correction factor to the rotational hydrodynamic mobility given by Eq.~\eqref{k2} versus the scaled radius of the coaxially positioned rigid disks for various values of the singularity position within the gap between the two disks.
The correction factor increases monotonically upon increasing the size of the disks because the rotational motion of the confined particle becomes more restricted.
In the limit of infinitely large disks, the correction factor asymptotically tends to the value given by Eq.~\eqref{k2inf}.

\subsubsection{Superposition approximation}

In the presence of two sufficiently separated confining boundaries, the correction to the hydrodynamic mobility can sometimes be approximated by superimposing the individual contributions arising from each boundary~\cite{dufresne01, benesch2003brownian, polin2006anomalous, daddi16b, mathijssen16jfm, daddi18jpcm, daddi2021hydrodynamics}.
This superposition approximation has originally been introduced by Oseen~\cite{oseen27} to estimate the translational hydrodynamic mobility in a channel bounded by two plates.
In this approach, the scaled correction factor can be approximated by
\begin{equation}
	k_2^\mathrm{Sup} = R^{-3}
	\left(  \xi_{-}^{-3} k_1 \left( \xi_{-} \right)
	+ \xi_{+}^{-3} k_1 \left( \xi_{+} \right) \right) , 
	\label{k2Sup}
\end{equation}
wherein~$\xi_\pm = \left( \frac{1}{2} \pm h \right)/R$ with $k_1$ representing the scaled correction factor near a single disk given by Eq.~\eqref{k1Final}.
In particular, near the upper wall, a Taylor expansion around $h = 1/2$ leads to
\begin{equation}
	k_2^\mathrm{Sup} = \frac{1}{8} \, \epsilon^{-3}
	+ k_1 \left( R^{-1} \right) + \mathcal{O} \left( \epsilon \right) \, .
\end{equation}

\begin{figure}
	\centering
	\includegraphics[scale=1]{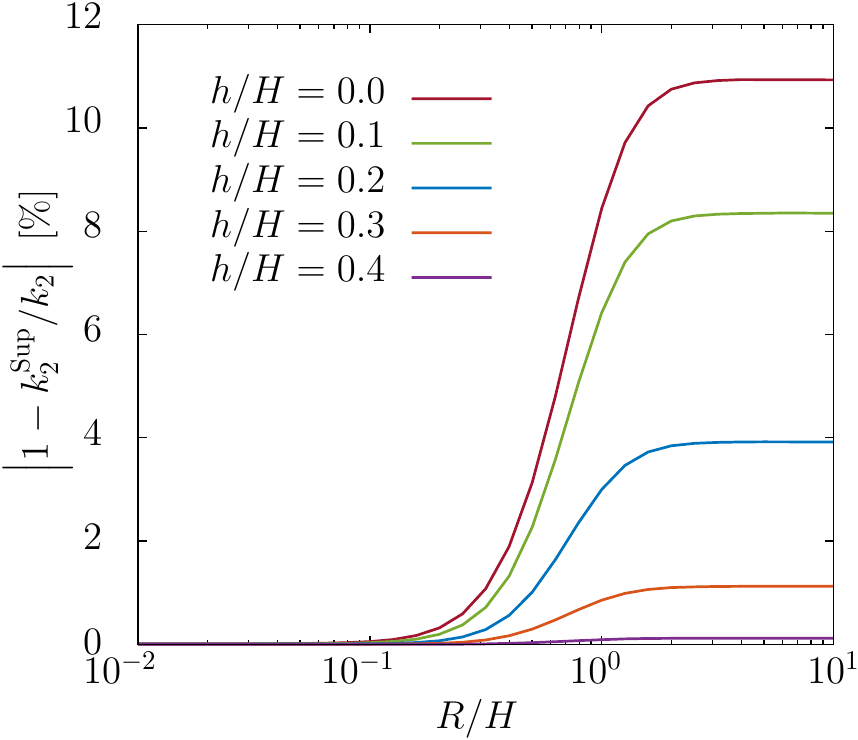}
	\caption{(Color online) Percentage relative error of the scaled correction factor of the rotational hydrodynamic mobility between two coaxially positioned rigid no-slip disks as obtained using the simplistic superposition approximation stated by Eq.~\eqref{k2Sup} in comparison to the exact formula given by Eq.~\eqref{k2}.}
	\label{ErrorSup}
\end{figure}

Figure~\ref{ErrorSup} displays the evolution of the percentage relative error committed by using Oseen's superposition approximation.
We observe that the error increases monotonically with the system size and reaches its maximum value in the limit $R \to \infty$.
In particular, the error attains its extreme value in the mid-plane of the channel for $h = 0$, where the MPRE amounts to about 11\%.
The latter value is remarkably lower than the one previously obtained for axisymmetric translational motion along the axis of two coaxially positioned rigid no-slip disks~\cite{daddi2020axisymmetric} for which the MPRE was found to be as high as 55\% in the mid-plane.
Consequently, the superposition approximation can be employed to estimate the rotational mobility between two confining disks without significantly compromising the accuracy of the prediction.

\section{Conclusions}
\label{sec:conclusions}

To summarize, we have presented an analytical and semi-analytical theory to quantify the low-Reynolds-number flow induced by a point-torque singularity acting near a single disk or between two coaxially positioned rigid disks, respectively, satisfying no-slip boundary conditions on the surfaces of the disks.
The rotlet is assumed to be located on the axis of symmetry of the disks with the torque directed along that axis.
We have formulated the solution of the hydrodynamic equations as mixed-boundary-value problems, which we subsequently reduced into systems of dual integral equations with Bessel-function kernels.
On the one hand, we have demonstrated that, near a single disk, the resulting integral equation can appropriately be transformed into a classical Abel integral equation that admits a unique solution.
On the other hand, we have shown that, between two coaxially positioned disks, a system of two Fredholm integral equations of the first kind arises.
For its solution, we have approximated the integral by standard middle Riemann sums reducing the dual integral equations into a linear system of equations amenable to immediate inversion using standard numerical approaches.

Moreover, we have made use of the derived solution of the flow problems to probe the effect of confinement on the rotational hydrodynamic mobility of a small colloidal particle through which the torque is exerted on the fluid.
More importantly, we have assessed the accuracy and reliability of Oseen's superposition approximation, which is commonly employed to predict the rotational mobility in confined geometries.
We have found that the maximum percentage relative error of Oseen's approximation is only about 11\% in the mid-plane of the channel, suggesting that this simplistic approximation could generally be employed to estimate the rotational mobility between two finite-sized disks.

The systems addressed in the present study may find useful applications in various biologically and technologically relevant processes.
On the one hand, the solution of the flow problem for a rotlet singularity acting near a finite-sized disk may be employed in the context of micromixing, as a small-scale analog to a magnetic stir bar mixer driven by an external rotating magnetic field.
On the other hand, the solution inside a gap bounded by two coaxially positioned disks may prove to be useful, for instance, in the modeling of ionic transport in small-scale capacitors.

The present results may be extended to further explore the rotational motion of a spherical particle of finite size, with a radius comparable to the radii of the confining disks.
For that purpose, the solution of the flow problem could, in principle, be formulated in bipolar coordinates.
Another possible extension of the present work could be to address the general problem of rotational motion near one or two no-slip disks for arbitrary positioning of the singularity and arbitrary orientation of the torque.
These steps could be the subject of possible future investigations.

\appendix

\section{Solution of the integral equation~\eqref{Abel}}
\label{appendixAbel}

In this Appendix, we show that the solution of the resulting integral equation Eq.~\eqref{Abel} can be cast in the form of solution of a classical Abel integral equation.
The general form of an Abel integral equation can be presented as~\cite{hochstadt2011integral} 
\begin{equation}
\int_0^x \frac{\phi(s)}{\left( x-s \right)^\frac{1}{2}} \, \Intd s = g(x) \, , \label{realAbel}
\end{equation}
the solution of which is given by
\begin{equation}
\phi(x) = \frac{1}{\pi} \frac{\Intd}{\Intd x} \int_0^x \frac{g(u)}{\left( x-u \right)^\frac{1}{2}} \, \Intd u \, . \label{soli}
\end{equation}

Using the changes of variables $s = t^2$ and $x = r^2$, Eq.~\eqref{Abel} can be written as
\begin{equation}
\int_0^{x} \frac{s^{-\frac{1}{4}} \, \hat{\omega} (s^\frac{1}{2})}{\left( x-s \right)^\frac{1}{2}} \, \Intd s = \left( 2\pi \right)^\frac{1}{2} x^\frac{1}{2} f(x^\frac{1}{2}) \, .
\end{equation}
By identification with Eq.~\eqref{realAbel}, we get
$\phi(s) = s^{-\frac{1}{4}} \, \hat{\omega} (s^\frac{1}{2})$ and 
$g(x) = \left( 2\pi \right)^\frac{1}{2} x^\frac{1}{2} f(x^\frac{1}{2})$.
Using the solution form given by Eq.~\eqref{soli}, we then obtain
\begin{equation}
x^{-\frac{1}{4}} \hat{\omega} (x^\frac{1}{2}) 
= \left( \frac{2}{\pi} \right)^\frac{1}{2} \frac{\Intd}{\Intd x} 
\int_0^x \frac{u^\frac{1}{2} f (u^\frac{1}{2})}{\left( x-u \right)^\frac{1}{2}} \, \Intd u \, .
\end{equation}

Next, applying the change of variable $r = x^\frac{1}{2}$, the latter equation can readily be expressed as
\begin{equation}
\hat{\omega} (r) = \frac{1}{2} \left( \frac{2}{\pi r} \right)^\frac{1}{2}
\frac{\Intd}{\Intd r} \int_0^{r^2} \frac{u^\frac{1}{2} f (u^\frac{1}{2})}{\left( r^2-u \right)^\frac{1}{2}}\, \Intd u \, .
\end{equation}

Finally, the change of variable $v = u^\frac{1}{2}$ yields
\begin{equation}
\hat{\omega} (r) = \left( \frac{2}{\pi r} \right)^\frac{1}{2} \frac{\Intd}{\Intd r} \int_0^r
\frac{v^2 f(v)}{\left( r^2-v^2 \right)^\frac{1}{2}} \, \Intd v \, , 
\end{equation}
which exactly corresponds to Eq.~\eqref{AbelInter} rewriting $r=t$.

\section{Comparison with numerical calculations using the finite-element method}
\label{appendixRichter}

To confirm our analytical solutions, we perform numerical simulations using the finite-element method. 
Formulated in cylindrical coordinates, $\vect{v}=(v_r,v_\theta,v_z)$, we can take
advantage of the fact that the solution is constant in angular
direction, $\partial_\theta \vect{v}=\vect{0}$. This reduces the problem to
a two-dimensional equation 
formulated in the $r/z$-plane for the angular component $v_\theta$
only. 
Since there is also no coupling to the pressure field, the problem reduces to a scalar one. 
In the following, we illustrate our approach for the two-disk geometry and denote by
\begin{equation}
\begin{split}
\Omega &= \Big((0,R_{max})\times (-Z_{max},Z_{max})\Big) \\
&\qquad\setminus
\Big( (0,R)\times \{H/2\} \cup (0,R)\times \{-H/2\}\Big)
\end{split}
\end{equation}
the numerical domain, artificially restricted to $0<r<R_{max}$ and
$-Z_{max}<z<Z_{max}$. To limit the impact of the artificial outer
boundaries, we set $R_{max}=Z_{max}=6.625$. We could not identify a significant
effect by further extending these limits.

The variational formulation of the problem is then given by \cite{Wahl2021}
\begin{equation} \label{F_phi}
\int_\Omega \eta r  \Big(
\frac{\partial v_\theta}{\partial r}  \frac{\partial \phi}{\partial r}
+\frac{\partial v_\theta}{\partial z}  \frac{\partial \phi}{\partial z}
+\frac{1}{r} v_\theta \phi
\Big)\,\text{d}r\,\text{d}z = F(\phi) \, ,
\end{equation}
$\forall \phi\in H^1_0(\Omega;D)$, where we denote by $H^1_0(\Omega;D)$ the space of square integrable
functions with weak derivatives that are zero on the two discs
$D=(0,R)\times \{-H/2\}\cup (0,R)\times \{H/2\}$. On the right-hand side of Eq.~\eqref{F_phi},
the problem is driven by a Dirac form as
\begin{equation}
F(\phi) = \phi(\vect{r}_h) \, ,
\end{equation}
centered in a point close to the $z$-axis $\vect{r}_h=(r_0,z_h)$,
$z_h=1/128$. 

The equation is discretized with quadratic finite-elements using
piecewise quadratic elements on a quadrilateral
mesh~\cite{Richter2017} featuring about $1\,750\,000$
unknowns. All computations are performed in the finite-element software
library Gascoigne 3d~\cite{Gascoigne}.

\section{Mobility between two disks in the limit $R \ll 1$}
\label{appendixMobi}

In the range of small values of~$R \ll 1$, we attempt to find an approximate expression of the scaled correction factor.
For $t \ll 1$, it follows from Eqs.~\eqref{Km} and \eqref{S} that $\mathcal{K}_{-}(t) \sim t^{\frac{1}{2}}$ and $\mathcal{S} (t) \sim t^{-\frac{1}{2}}$, respectively.
Therefore, to ensure the convergence of the system of integral equations~\eqref{SystDualIntEqns} at the lower limit $t=0$, we require that the unknown functions $\hat{f}_1(t)$ and~$\hat{f}_2(t)$ scale (at least) as $t^{\frac{1}{2}}$ as $t\to 0$.

We now use the ansatz $\hat{f}_1(t) = \alpha_1 t^\frac{1}{2}$ and~$\hat{f}_2(t) = \alpha_2 t^\frac{1}{2}$, where $\alpha_1$ and~$\alpha_2$ are two real numbers to be subsequently determined.
Inserting these expressions into Eqs.~\eqref{SystDualIntEqns} evaluated at $r = \beta R$, with $\beta \in (0,1)$, performing analytically the integration, expanding the resulting expressions into Taylor series of $R$, and solving for $\alpha_1$ and $\alpha_2$ yields
\begin{equation}
\alpha_1 = -\left( \frac{\pi}{2} \right)^\frac{1}{2} \frac{\beta R}{\left( \frac{1}{2}+h \right)^3} \, , \quad
\alpha_2 = \left( \frac{\pi}{2} \right)^\frac{1}{2} \frac{\beta R}{\left( \frac{1}{2}-h \right)^3} \, .	
\label{alpha1UNDalpha2}
\end{equation}

Next, by substituting the above expressions of $\hat{f}_1(t)$ and~$\hat{f}_2(t)$ into Eq.~\eqref{k2}, evaluating the integral analytically and performing a series expansion about $R=0$, we obtain
\begin{equation}
k_2 \simeq \frac{R^2}{2} \left( \frac{2}{\pi} \right)^\frac{1}{2} 
\left( \frac{\alpha_2}{\left( \frac{1}{2} - h \right)^3} 
- \frac{\alpha_1}{\left( \frac{1}{2} + h \right)^3} \right) \, .
\label{GreatAndreas}
\end{equation}

Finally, by inserting the expressions of~$\alpha_1$ and~$\alpha_2$ stated by Eqs.~\eqref{alpha1UNDalpha2} into Eq.~\eqref{GreatAndreas}, the correction factor in the range $R \ll 1$ can be obtained as
\begin{equation}
k_2 \simeq
\frac{\beta  R^3}{2} \left( \left( \frac{1}{2} - h \right)^{-6} 
+ \left( \frac{1}{2} + h \right)^{-6} \right) \, .
\end{equation}
By setting $\beta = 5/6$, the latter approximate expression is found to be in good agreement with the numerical results.

\section*{Acknowledgments}

A.D.M.I.\ and H.L. gratefully acknowledges support from the DFG (Deutsche Forschungsgemeinschaft) through the projects DA~2107/1-1 and LO~418/23-1. 
A.M.M.\ thanks the DFG for support through the Heisenberg Grant ME~3571/4-1.

%


\end{document}